# Study of scaling in the fractional quantum Hall effect

X.C. Xie[1], D.Z. Liu[2], and J.K. Jain[3]

[1] Department of Physics, Oklahoma State University, Stillwater, OK 74078.
[2] James Franck Institute, University of Chicago, Chicago, IL 60637
[3] Department of Physics, State University of New York at Stony Brook, Stony Brook, New York 11794-3800

(Draft on December 4, 1995)

In the composite fermion model of the fractional quantum Hall effect, composite fermions experience, in addition to the usual potential disorder, also a magnetic flux disorder. Motivated by this, we investigate the localization properties of a single fermion in two dimensions, moving in the presence of both potential and magnetic flux disorders, but with a non-zero average magnetic field. It is found that the exponent characterizing the divergence of the localization length is not changed upon the addition of the flux disorder, provided it is not too large, suggesting that the transitions between fractionally quantized Hall plateaus belong to the same universality class as those between the integrally quantized ones.

PACS numbers: 73.40.Hm

In the absence of a magnetic field, an arbitrary weak random potential localizes all states of an electron in two dimensions [1]. In contrast, existence of extended states, separated by bands of localized states, is crucial for the explanation of the quantum Hall effect [2]. Numerical studies [3,4] have made a convincing case that the localization length of an electron, confined to two dimensions and exposed to a strong magnetic field, diverges at a single energy near the center of each Landau level, and the divergence is characterized by an exponent $\alpha$:

$$\xi = |E - E_c|^{-\alpha} , \qquad (1)$$

where $\alpha \approx 7/3$, independent of the Landau level (LL) index. No analytical derivation of the value of the exponent yet exists. Experimentally, the width of the transition region (e.g., the width of the conductance peak) in the integer quantum Hall regime (IQHE) regime has been found to vanish in a range of temperature as $T^\beta$, where $\beta \approx 0.42$, independent of the LL in question [5]. This is consistent with the theoretical value of $\alpha$ in a variable range hopping theory [6], which obtains $\beta = 1/\alpha$ (in other words, writing $\beta = 1/z\alpha$, this work shows that the dynamical exponent $z$ is equal to unity for the Coulomb interaction). Direct measurements of the localization length are also in agreement with $\alpha \approx 7/3$ [7].

At first glance, the nature of localization would appear to be much more complex in the fractional quantum Hall effect (FQHE) regime, since interactions are an intimate part of the physics here and cannot be neglected. However, in the composite fermion model [8], the essential part of the interelectron interaction is incorporated through the formation of composite fermions, which themselves can be treated as non-interacting to a good approximation. An effective single-particle description of the strongly correlated electron state thus becomes possible in terms of composite fermions. This paper studies the scaling of the localization length in the FQHE regime in the framework of a non-interacting composite fermion model. While the dynamics of composite fermions is similar to that of electrons in a weaker magnetic field, an essential difference is that the composite fermions also experience a random flux disorder in addition to the usual random potential disorder [9–11]. The main conclusion of the paper is that the exponent $\alpha$ is the same for non-interacting composite fermions as for non-interacting electrons, provided the amplitude of the random flux disorder is smaller than the average magnetic field. This is consistent, following Ref. [6], with the experimental result that the exponent for the transition from 1/3 to 2/5 is the same as for the transitions between IQHE states [12].

The result is not entirely unexpected, and earlier works have also anticipated that the FQHE transitions belong to the same universality class as the IQHE transitions. Jain, Kivelson and Trivedi [13] argued that the composite fermion (CF) wave functions describe the physics even in the presence of disorder; this leads to a law of corresponding states relating the FQHE states to the IQHE states, and makes it plausible that the localization length of composite fermions is the same as that of electrons to which their wave function is related, leading to the same exponent. The validity of the CF wave functions in the presence of disorder was demonstrated only in some limited cases, however. The present work investigates the problem from a different point of view, and sheds further light on the issue.

According to the CF *ansatz*, in a range of filling factors in the lowest LL, electrons find it energetically favorable to capture an even number of vortices of the many particle wave function. The bound state of an electron and vortices behaves as a particle, called the composite fermion. The vortices produce phases as the composite fermions move around, which partly cancel the Aharonov-Bohm phases originating from the external magnetic field, and, as a result, the composite fermions experience an effective magnetic field given by

$$B^* = B - 2m\rho\phi_0 ,  \quad (2)$$

where $B$ is the external field, $2m$ is the number of vortices carried by composite fermions, $\phi_0 = hc/e$ is the magnetic flux quantum, and $\rho$ is the electron (or CF) density. In this manner, the problem of strongly correlated electrons at $B$ is mapped on to the problem of weakly interacting composite fermions at $B^*$ and the FQHE transitions of electrons are mapped onto the IQHE transitions of composite fermions. One might be tempted immediately to conclude that the same exponent $\alpha$ should characterize the divergence of the localization length. However, there is one important difference between the IQHE transitions of electrons and composite fermions: while for electrons the magnetic field is constant, the effective magnetic field for the composite fermions fluctuates in space, since a potential disorder produces density variations, which, through the above equation, leads to fluctuations in the magnetic field experienced by the composite fermion [9,10]. Motivated by this picture, we consider a single composite fermion in an average magnetic field $B^*$, in the presence of a random potential as well as a (static) random flux disorder.

In the following, we briefly outline our model and technique. Our numerical sample consists of a very long two-dimensional strip of a finite width ($M$) square lattice with nearest neighbor hopping. Periodic boundary condition in the width direction is used to avoid complications arising from the edge states. The disorder potential is modeled by an on-site white-noise potential $V_{im}$ ($i$ denotes the column index, $m$ denotes the chain index) ranging from $-W/2$ to $W/2$. The magnetic field appears through the complex phase of the hopping term. The strength of the magnetic field is characterized by the flux per plaquette ($\phi$) measured in unit of $\phi_0$. It is the sum of an average field $\phi_a$ and a random field distributed uniformly between $-\phi_r/2$ to $+\phi_r/2$. The Hamiltonian of this system can be written as:

$$\mathcal{H} = \sum_i \sum_{m=1}^M V_{im} |im><im| \quad (3)$$
$$+ \sum_{<im;jn>} \left[ t_{im;jn} |im><jn| + t^\dagger_{im;jn} |jn><im| \right],$$

where $<im;jn>$ indicates nearest neighbors on the lattice. The amplitude of the hopping term is chosen as the unit of energy. The gauge is conveniently chosen so that the inter-column hopping does not carry a complex phase factor (i.e., $t_{im;i+1,m} = -1$) and the effect of magnetic field shows up only through the phase factor of the intra-column (inter-chain) hopping term. If the random flux in a plaquette cornered by $(im), (i+1,m), (i+1,m+1)$ and $(i,m+1)$ is $\phi_{im}$, then

$$\frac{t_{i+1,m;i+1,m+1}}{t_{im;i,m+1}} = \exp\left[i2\pi \frac{\phi_{im}}{\phi_o}\right]. \quad (4)$$

For a specific energy $E$, a transfer matrix $T_i$ can be easily set up mapping the wavefunction amplitudes at columns $i-1$ and $i$ to those in columns $i$ and $i+1$, i.e.

$$\begin{pmatrix} \psi_{i+1} \\ \psi_i \end{pmatrix} = T_i \begin{pmatrix} \psi_i \\ \psi_{i-1} \end{pmatrix} = \begin{pmatrix} H_i - E & -I \\ I & 0 \end{pmatrix} \begin{pmatrix} \psi_i \\ \psi_{i-1} \end{pmatrix}, \quad (5)$$

where $H_i$ is the Hamiltonian for the $i$th column and $I$ is an $M \times M$ unit matrix. Using a standard iteration algorithm [4], we can calculate the Lyapunov exponents for the transfer matrix $T_i$. The finite-size localization length $\lambda_M(E)$ at energy $E$ for finite width $M$ is then given by the inverse of the smallest Lyapunov exponent. In our numerical calculation, we choose the sample length to be over $10^4$ so that the self-averaging effect automatically takes care of the ensemble statistical fluctuations. We then obtain the thermodynamic localization length $\xi$ using the standard one-parameter finite-size scaling analysis [14], according to which the renormalized finite-size localization length $\lambda_M/M$ can be expressed in terms of a universal function of $M/\xi$, i.e.,

$$\frac{\lambda_M(E)}{M} = f\left(\frac{M}{\xi(E)}\right), \quad (6)$$

where $f(x) \propto 1/x$ in the thermodynamic limit ($M \to \infty$) for localized states while approaching a constant ($\sim 1$) when $\xi$ diverges. The critical exponent $\alpha$, given in Table I, is obtained from the best fit of $\xi = |E - E_c|^{-\alpha}$. The value of $E_c$ is obtained using the method of Ref. [15].

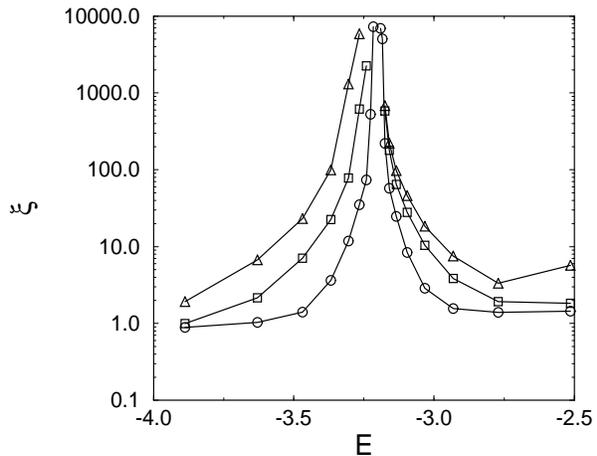

FIG. 1. Thermodynamic localization length $\xi$ for average field $\phi_a = 1/7$ and zero random field $\phi_r = 0$ with disorder $W = 1.0$ (circles), $W = 2.0$ (squares) and $W = 3.0$ (triangles).

We focus below on the localization behavior in the lowest Landau level, i.e. on the $1 \to 2$ transition of composite fermions, which corresponds to the $1/3 \to 2/5$ transition of electrons. Throughout this work, the average field is fixed at $\phi_a = 1/7$. Fig. 1 presents the results for the disorder dependence of the thermodynamic localization length $\xi$ with zero random field. The results are clearly consistent with the notion that all states are localized except at a single energy, $E_c$, which is determined to be $E_c \simeq -3.2$ from the method of Ref. [15]. We note that $E_c$ can also be obtained from numerical fitting, which gives almost identical answer. The main effect of the disorder is to change the localization length; interestingly, in the present problem, increasing $W$ causes the $\xi$ to increase in contrast to the usual intuition. [16]

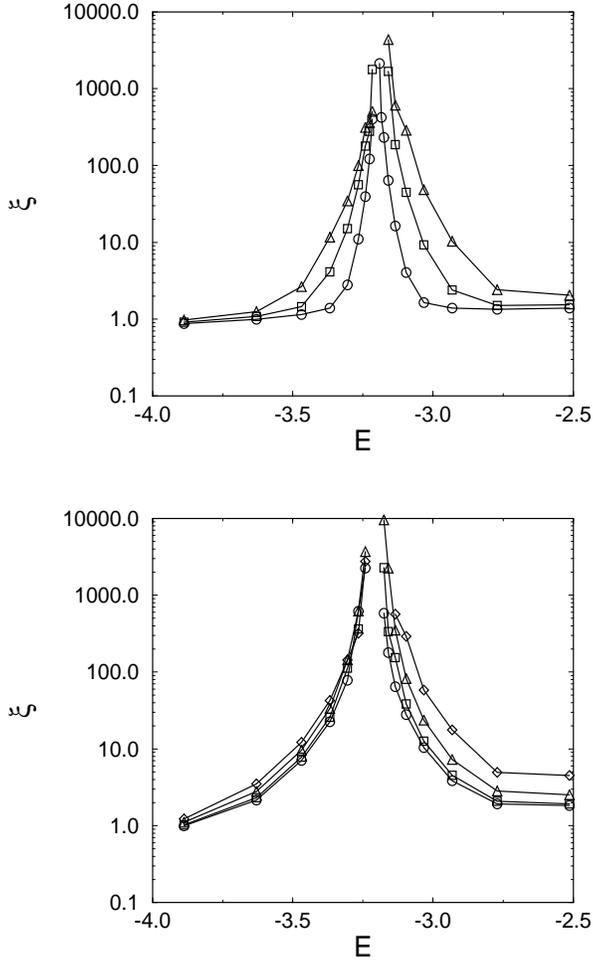

FIG. 2. Thermodynamic localization length $\xi$ for average field $\phi_a = 1/7$. (a) $W = 0$ and random fields $\phi_r = 0.1$ (circles), $\phi_r = 0.2$ (squares) and $\phi_r = 0.3$ (triangles); (b) $W = 2$ and random fields $\phi_r = 0$ (circles), $\phi_r = 0.1$ (squares), $\phi_r = 0.2$ (triangles) and $\phi_r = 0.3$ (diamonds).

Next we compute the localization length $\xi$ as a function of the random magnetic field (for several values of disorder), shown in Fig. 2. For a given energy, $\xi$ increases for increasing random field $\phi_r$, indicating that the role of the random field is qualitatively similar to that of the random disorder. The critical exponent $\alpha$ is again calculated by fitting $\xi(E) \propto |E - E_c|^{-\alpha}$, and listed in Table I below. Two exponents are given for each value of $W$ and $\phi_r$, one obtained from the lower energy side and the other from the higher energy side. The error shown in Table I for each critical exponent represents the error of the least square fit for $\xi$ versus energy plot. Another contribution to the uncertainty, of order $\sim 0.3$, comes from the various possible choices of points close to $E_c$ used for the fit. The exponents are consistent with 7/3 within our numerical uncertainty, demonstrating that the addition of random magnetic field does not alter the universality of the transition.

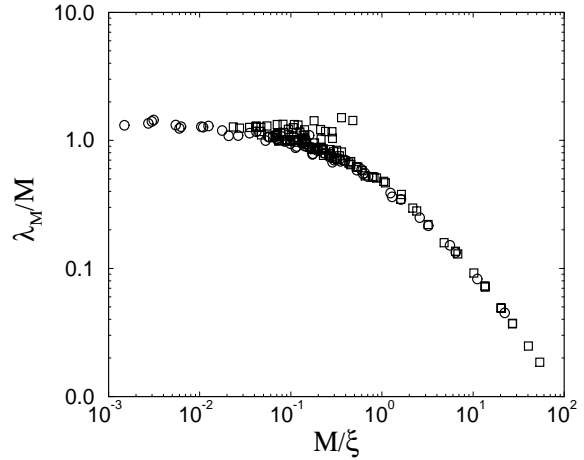

FIG. 3. Scaling function for average field $\phi_a = 1/7$ and disorder $W = 1.0$ with random field $\phi_r = 0$ (circles) and $\phi_r = 0.6$ (squares).

Until now, we have restricted our study to the situation where the amplitude of the random field is smaller than the average field ($\phi_r = 0.3$ being a borderline case, with $\phi_r/2$ only slightly larger than $\phi_a = 1/7$). Under this condition, the chirality of the particle does not change in the sense that the cyclotron motion is everywhere in the same direction. Once the amplitude of the random field exceeds the average field, the cyclotron motion can be clockwise or counter-clockwise in various regions of the two dimensional system. Fig. 3 plots the scaling function, i.e., $\lambda_M/M$ versus $M/\xi$ for $\phi_a = 1/7$ and $W = 1$ with $\phi_r = 0$ (circles) and $\phi_r = 0.6$ (squares). The curve with circles is a typical representation of the scaling function for all the cases listed in Table I. In all those cases, we found the single-parameter scaling hypothesis of Eq.(6) is satisfied in the sense that all data fall into one smoothly universal curve similar to the circles in Fig.3. On the other hand, the data (squares) for $\phi_r = 0.6$ clearly do not fall into one smooth curve, indicating a lack of scaling behavior. In particular, there are no data points for $M/\xi \to 0$, suggesting that $\xi$ does not diverges. In Fig. 4 we plot $\xi$ for $W = 1$ (circles), $W = 2$ (squares) and $W = 3$ (triangles). The result is qualitatively different from that in Fig. 2, especially on the high energy side of $E_c$, where the CF eigenstates are much more extended, and the localization length fluctuates strongly as a function of the energy or disorder. Further study is needed to clarify several issues here.

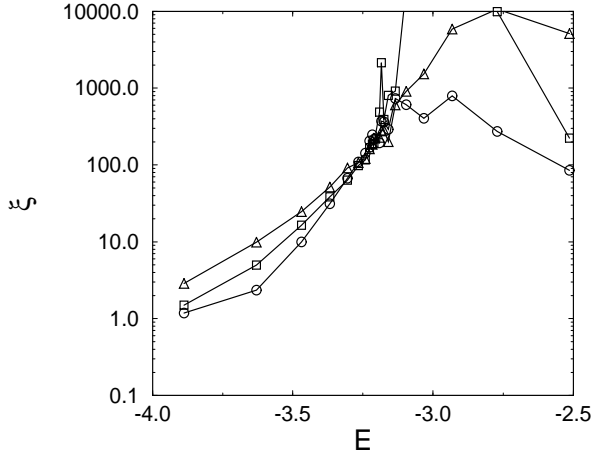

FIG. 4. Thermodynamic localization length $\xi$ for average field $\phi_a = 1/7$ and random field $\phi_r = 0.6$ with disorder $W = 1.0$ (circles), $W = 2.0$ (squares) and $W = 3.0$ (triangles).

Before closing, we discuss an important assumption in the above model, namely that of the neglect of the residual interaction between composite fermions. In addition to potential and static flux disorders, the inter-CF interaction will also produce dynamical flux disorder. However, various theoretical and experimental studies have shown that the interaction between the composite fermions is weak, and can be included perturbatively. Studies treating the Coulomb interaction perturbatively in the IQHE regime have found that it does not affect the critical exponent [17]. This suggests that the inter-CF interaction is also unlikely to alter the critical behavior.

This work was supported in part by the NSF under Grant no. DMR91-20000(DZL) and DMR93-18739 (JKJ). We are grateful to X.G. Wen for a discussion.

TABLE I. The critical exponent($\alpha$) for several values of random magnetic field ($\phi_r$) and on-site disorder ($W$).

| $W, \phi_r$ | 0.0 | 0.1 | 0.2 | 0.3 |
|---|---|---|---|---|
| 0 |  | $2.50 \pm 0.064$ | $2.67 \pm 0.27$ | $2.84 \pm 0.7$ |
|  |  | $2.48 \pm 0.069$ | $2.30 \pm 0.27$ | $2.17 \pm 0.57$ |
| 1 | $2.34 \pm 0.48$ | $2.19 \pm 0.30$ | $2.53 \pm 0.13$ | $2.32 \pm 0.54$ |
|  | $2.20 \pm 0.26$ | $2.54 \pm 0.27$ | $2.40 \pm 0.088$ | $2.21 \pm 0.098$ |
| 2 | $2.03 \pm 0.083$ | $2.44 \pm 0.16$ | $2.57 \pm 0.10$ | $2.61 \pm 0.083$ |
|  | $2.68 \pm 0.43$ | $2.56 \pm 0.28$ | $2.60 \pm 0.22$ | $2.44 \pm 0.066$ |
| 3 | $1.85 \pm 0.029$ | $1.82 \pm 0.027$ | $2.53 \pm 0.23$ | $2.19 \pm 0.052$ |
|  | $2.78 \pm 0.66$ | $2.79 \pm 0.14$ | $2.65 \pm 0.11$ | $2.88 \pm 0.14$ |